\documentclass[12pt,aps,prb,showpacs,preprint]{revtex4}
\usepackage{amsmath}
\usepackage{amssymb}
\usepackage{graphicx}
\begin{document}
\title{APPROXIMATION TECHNIQUES FOR NON LINEAR OSCILLATORS}
\author{J.K. Bhattacharjee}
\email{jkb@bose.res.in}
\affiliation{S.N. Bose National Centre for Basic Sciences, Salt Lake, Kolkata-700098, India}
\author{Debabrata Dutta}
\email{intermittency@gmail.com}
\affiliation{S.N. Bose National Centre for Basic Sciences, Salt Lake, Kolkata-700098, India}
\author{Amartya Sarkar}
\email{amarta345@bose.res.in}
\affiliation{S.N. Bose National Centre for Basic Sciences, Salt Lake, Kolkata-700098, India}
\begin{abstract}
We show that the Lindstedt-Poincare perturbation theory is always a reliable technique in the region of small coupling constant. The harmonic balance result, on the other hand, if expanded in the perturbation parameter may lead to incorrect results.
\end{abstract}

\maketitle
\section{INTRODUCTION}
A senior undergraduate course in nonlinear dynamics\cite{nldstr,clmlanlif} abounds in various kinds of approximation procedures. There is the Lindstedt Poincare method, the harmonic balance, the multiple time scale analysis, Bogoluibov-Krylov technique and so on. It is interesting to note that for the cubic anharmonic oscillator the Lindstedt-Poincare perturbation theory answer given in the text by Landau and Lifshitz differs from the harmonic balance answer of Jordan and Smith if the Jordan and Smith result is expanded in a perturbation series in the cubic nonlinearity. In view of this difference we re-examine the Lindstedt-Poincare and the harmonic balance technique and compare with the perturbative expansion of the exact integral for the frequency of the motion.The Lindstedt-Poincare expansion is in agreement with the perturbative expansion of the integral. We point out how the harmonic balance method can be combined with perturbation theory to yield an answer in agreement with the Lindstedt-Poincare expansion. As another example of a possible discrepancy, we consider the periodic orbit of the Lotka-Volterra model. Application of the Lindstedt-Poincare technique yields an answer which differs from the perturbation expansion of the harmonic balance result given in Jordan and Smith. A numerical determination of the frequency shows agreement with the Lindstedt-Poincare method. It is apparent that the Lindstedt-Poincare method is always reliable. A perturbative expansion of a non-systematic harmonic balance answer may not give the right perturbation series when the equation of motion does not have reflection symmetry. To get the low coupling constant part right, harmonic balance has to be used in conjunction with perturbation theory. In case there is reflection symmetry, the frequency, determined by harmonic balance is automatically in agreement with perturbation theory. We point out that knowing the correct perturbation theory is vital to constructing a globally valid formula for the frequency in an equivalent linearisation technique.


\section{EXACT INTEGRAL AND LINDSTEDT-POINCARE PERTURBATION THEORY}
We use the example of the anharmonic oscillator
\begin{equation}\label{p1,eqn1}
\ddot{x}+\omega^{2}x+\lambda x^{3}=0
\end{equation}
to illustrate Lindstedt-Poincare, harmonic balance and the equivalent linearization techniques. This system is in principle exactly solvable as we can write the integral of motions as
\begin{equation}\label{p1,eqn2}
\frac{1}{2}\left(\frac{dx}{dt}\right)^2+\frac{1}{2}\omega^2 x^2+\frac{\lambda}{4} x^4 = constant
\end{equation}
The constant can be evaluated by noting that $\dot{x}=0$ at the turning point i.e. at $x=a$, where $a$ is the amplitude of motion. This leads to 
\begin{equation}\label{p1,eqn3}
\left( \frac{dx}{dt}\right)^2 = \omega^2 \left( a^2 - x^2\right) + \frac{\lambda}{2}\left(a^4 - x^4 \right)
\end{equation}
and consequently
\begin{equation}\label{p1,eqn4}
 dt=\dfrac{dx}{\left[ \omega^2 \left( a^2 - x^2\right) + \frac{\lambda}{2}\left(a^4 - x^4 \right)\right]^{1/2}}
\end{equation}
For the symmetric potential, that we have here $V(x)=\frac{1}{2}\omega^2 x^2+\frac{\lambda}{4} x^4$, the time period $T$ is found from Eq.~\eqref{p1,eqn4} as
\begin{eqnarray}
\frac{T}{4} &=& \int_0^a{\frac{dx}{\omega\left(a^2-x^2\right)^{1/2}\left[1+\frac{\lambda}{2\omega^2}\left( 	a^2+x^2\right)\right]^{1/2}}}\nonumber\\
\phantom{\frac{T}{4}}&=& \int_{0}^{\pi/2}\frac{d\theta}{\omega\left[1+\frac{\lambda a^2}{2\omega^2}\left(1+\sin^2{\theta} \right)\right]}\phantom{uuuuuuuu}\label{p1,eqn5}
\end{eqnarray}
We can express the right hand side of the above equation in terms of the elliptic function. For $\lambda a^2/2\omega^2\ll 1$, we can expand the integrand as,
\begin{eqnarray}
\frac{T}{4}&=&\int_0^{\pi/2}{\frac{d\theta}{\omega}}-\int_0^{\pi/2}{\frac{d\theta}{\omega}}\left(1+\sin^2\theta\right)+\dots\nonumber\\
&=&\frac{\pi}{2\omega}\left[ 1-\frac{3\lambda a^2}{8\omega^3}\right]\phantom{uuuuuuuuuuuuuuuuu}\label{p1,eqn6}
\end{eqnarray}
while for $\lambda a^2/\omega^2\gg1$
\begin{eqnarray}
\frac{T}{4} &\simeq & \frac{1}{\omega}\int_0^{\pi/2}\sqrt{\frac{2\omega^2}{\lambda a^2}}\frac{d\theta}{\left(1+\sin^2{\theta}\right)^{1/2}}\nonumber\\&=&\sqrt{\frac{4}{\lambda a^2}}\frac{1}{4}\Gamma(1/4)\phantom{uuuuuuuuuuu}\label{p1,eqn7}
\end{eqnarray}
The frequency $\Omega$ of the oscillations is $\Omega=2\pi/T$ and is found from Eqs.~\eqref{p1,eqn6} and ~\eqref{p1,eqn7} to be given by
\begin{equation}\label{p1,eqn8}
\Omega=\omega+\frac{3}{8}\frac{\lambda a^2}{\omega}+\dots\quad\quad\quad\quad\frac{\lambda a^2}{\omega^2}\ll 1
\end{equation}
and
\begin{equation}\label{p1,eqn9}
\Omega\simeq\pi\frac{{(\lambda a^2)}^{1/2}}{\Gamma\left( \frac{1}{4}\right) }\quad\quad\quad\quad\quad\quad\frac{\lambda a^2}{\omega^2}\gg 1
\end{equation}
The Lindstedt-Poincare perturbation theory for Eq.~\eqref{p1,eqn1} proceeds by expanding
\begin{equation}\label{p1,eqn10}
x=x_0+\lambda x_1+\lambda^2 x_2+\dots
\end{equation}
and writing the actual frequency $\Omega$ as 
\begin{equation}\label{p1,eqn11}
\Omega^2=\omega^2+\lambda\omega_1^2+\lambda^2\omega_2^2+\dots
\end{equation}
In terms of the above expansions, Eq.~\eqref{p1,eqn1} can be written as
\begin{eqnarray}
\ddot{x_0}+\lambda\ddot{x_1}+\lambda^2\ddot{x_2}+\dots+\Omega^2\left(x_0+\lambda x_1+\lambda^2 x_2+\dots\right)\phantom{uuuuuuuuuuuuuuuuuuu}\nonumber\\ = -\lambda\left(x_0+\lambda x_1+\dots\right)+\lambda\omega_1^2\left(x_0+\lambda x_1+\dots\right)+\lambda^2\omega_2^2\left(x_0+\lambda x_1+\dots\right)\label{p1,eqn12}
\end{eqnarray}
Equating the same powers of $\lambda$ on either sides, we get
\begin{subequations}
\begin{equation}\label{p1,eqn13a}
\lambda^0 \colon\quad\quad\ddot{x}_0+\Omega^2 x_0 = 0\phantom{uuuuuuuuuuuuuuuu}
\end{equation}
\begin{equation}\label{p1,eqn13b}
\lambda^1 \colon\quad\quad\ddot{x}_1+\Omega^2 x_1 = -x_0^3+\omega_1^2 x_0\phantom{uuuuuuuu}
\end{equation}
\begin{equation}\label{p1,eqn13c}
\lambda^2 \colon\quad\quad\ddot{x}_2+\Omega^2 x_2 = -3 x_0^2 x_1+\omega_1^2 x_1 + \omega_2^2 x_0
\end{equation}
\end{subequations}
The solution of Eq.~\eqref{p1,eqn13a} for the initial conditions $x_0=A_0,\dot{x_0}=0$ at $t=0$ is
\begin{equation}\label{p1,eqn14}
 x_0=A_0\cos{\Omega t}
\end{equation}
With $x_0$ obtained, Eq.~\eqref{p1,eqn13b} becomes
\begin{eqnarray}
\ddot{x}_1+\Omega^2 x_1 &=& -A_0^3\cos^3{\Omega t}+\omega_1^2 A_0\cos{\Omega t} \phantom{uuuuuuuuu}\nonumber\\
&=&\left(-\frac{3A_0^3}{4}+\omega_1^2 A_0\right)\cos{\Omega t}-\frac{A_0^3}{4}\cos{3\Omega t}\label{p1,eqn15}
\end{eqnarray}
The drive with frequency $\Omega$ on the right hand side causes a spurious resonance in the system. This would cause $x_1$ to diverge. In order to have a finite $x_1$, we need to remove the resonance causing term from the right hand side of Eq.~\eqref{p1,eqn15}. This is done by the choice
\begin{equation}\label{p1,eqn16}
\omega_1^2=\frac{3 A_0^2}{4}
\end{equation}
The equation for $x_1$ can now be solved and for the frequency $\Omega$ we have 
\begin{equation}\label{p1,eqn17}
\Omega^2 =\omega^2+\frac{3\lambda A_0^2}{4}+\dots
\end{equation}
or
\begin{equation}\label{p1,eqn18}
\Omega =\omega+\frac{3\lambda}{8}\frac{A_0^2}{\omega}+\dots
\end{equation}
which is the same as that shown in Eq.~\eqref{p1,eqn8}

Harmonic balance, on the other hand, requires the expansion (without loss of generality, we consider a solution with $\dot{x}=0)$ at $t=0$
\begin{equation}\label{p1,eqn19}
x=\alpha_0+\alpha_1\cos{\Omega t}+\alpha_2 \cos{2\Omega t}+\alpha_3 \cos{3\Omega t}+\dots
\end{equation}
Inserting this solution in Eq.~\eqref{p1,eqn1}
\begin{eqnarray}
-\Omega^2\alpha_1\cos{\Omega t}-4\Omega^2\alpha_2\cos{2\Omega t}-9\Omega^2\alpha_3\cos{3\Omega t}+\dots\phantom{uuuuuuuuuuuuuuuuuuuuuu}\nonumber\\+\omega^2\left(\alpha_0+\alpha_1\cos{\Omega t}+\alpha_2 \cos{2\Omega t}+\dots\right)\phantom{uuuuuuuuuuuuuuuuuuuuuuuu}\nonumber\\
+\lambda\left(\alpha_0^3+3\alpha_1\alpha_0^2\cos{\Omega t}+3\alpha_1^2\alpha_0\cos^2{\Omega t}+\alpha_1^3\cos^3{\Omega t}+\dots\right)=0\label{p1,eqn20}
\end{eqnarray}
Equating the coefficient of each harmonic separately to zero:
\begin{eqnarray}
\alpha_0 &=& 0 \\ \label{p1,eqn21}
\Omega^2 &=& \omega^2+\frac{3\alpha_1^2}{4}\lambda \label{p1,eqn22}
\end{eqnarray}
%


\section{THE ANHARMONIC OSCILLATOR}
We begin this section with the cubic oscillator
\begin{equation}\label{p1,eqn29}
\ddot{x}+\omega^2 x+\lambda x^2 = 0
\end{equation}
The potential is $V(x)=\frac{1}{2}\omega^2 x^2+\frac{\lambda}{3}x^3$ and we note that while it is unbounded for negative $x (\lambda >0)$, it has positive maximum at $x=-\omega^2/\lambda$ and if the total energy $E$ (determined by initial conditions) is such that it is less than $\frac{1}{6}\omega^4/\lambda$, then there will be periodic motion with frequency $\Omega$, with the time period given by
\begin{equation}\label{p1,eqn30}
T=\int_{a_1}^{a_2}\frac{dx}{\sqrt{2E-\omega^2 x^2-\frac{2\lambda x^3}{3}}}
\end{equation}
where $a_1$ and $a_2$ are the consecutive negative and positive real roots of the amplitude equations
\begin{eqnarray}
E=\frac{\omega^2 a^2}{2}+\frac{\lambda a^3}{3}\nonumber\\ 
\text{or}\quad\quad a^2=\frac{2E}{\omega^2}-\frac{2}{3}\frac{a^3\lambda}{\omega^2}\label{p1,eqn31}
\end{eqnarray}
For a perturbative expansion of the integral, we first need to find $a_1$ and $a_2$ from a perturbative determination of the roots of the cubic of Eq.~\eqref{p1,eqn31}. Expanding the root $A$ as
\begin{equation}\label{p1,eqn32}
A=A_0+\lambda A_1+\lambda^2 A_2+\dots
\end{equation}
we have from Eq.~\eqref{p1,eqn31}
\begin{equation}\label{p1,eqn33}
A_0^2+2\lambda A_1A_2+\lambda^2 A_1^2+2\lambda^2 A_0 A_2+\dots=\frac{2E}{\omega^2}-\frac{2}{3}\frac{\lambda}{\omega^2}\left(A_0^3+3\lambda A_0^2A_1+\dots\right)
\end{equation}
Equating equal powers of $\lambda$ on either sides,
\begin{subequations}
\begin{equation}\label{p1,eqn34a}
A_0^2=\frac{2E}{\omega^2}
\end{equation}
\begin{equation}\label{p1,eqn34b}
A_1 = -\frac{A_0^2}{3\omega^2}
\end{equation}
\begin{equation}\label{p1,eqn34c}
A_2 = \frac{5}{18}\frac{A_0^3}{\omega^4}
\end{equation}
\end{subequations}
At the zeroeth order $a_1=-\sqrt{\frac{2E}{\omega^2}}$ and $a_2=\sqrt{\frac{2E}{\omega^2}}$. To $\mathcal{O}(\lambda^2)$, we have from above
\begin{eqnarray}
a_1=\sqrt{\frac{2E}{\omega^2}}-\lambda\frac{2E}{3\omega^4}-\lambda^2\frac{5}{18}\frac{1}{\omega^4}\left(\frac{2E}{\omega^2}\right)^{3/2}+\dots\\ \label{p1,eqn35}
a_2=\sqrt{\frac{2E}{\omega^2}}-\lambda\frac{2E}{3\omega^4}+\lambda^2\frac{5}{18}\frac{1}{\omega^4}\left(\frac{2E}{\omega^2}\right)^{3/2}+\dots \label{p1,36}
\end{eqnarray}
Noting that $a_1<0$ and $a_2>0$, we can split the integral in Eq.~\eqref{p1,eqn30} as
\begin{eqnarray}
\frac{T}{2} &=& \int_{a_1}^0{\frac{dx}{\sqrt{2E-\omega^2 x^2- \frac{2\lambda x^3}{3}}}}+\int_0^{a_2}{\frac{dx}{\sqrt{2E-\omega^2 x^2-\frac{2\lambda x^3}{3}}}}\phantom{uuuuuuuuuuuuuuuuuuuuuuuuu}\nonumber\\
&=& \int_0^{|a_1|}{\frac{dx}{\sqrt{2E-\omega^2 x^2+\frac{2\lambda x^3}{3}}}}+\int_0^{a_2}{\frac{dx}{\sqrt{2E-\omega^2 x^2-\frac{2\lambda x^3}{3}}}}\phantom{uuuuuuuuuuuuuuuuuuuuuuuu}\nonumber\\ 
&=& \int_0^{|a_1|}{\frac{dx}{\sqrt{\omega^2\left(a_1^2-x^2\right)-\frac{2\lambda}{3}\left(|a_1|^3-x^3\right)}}}+\int_0^{a_2}{\frac{dx}{\sqrt{\omega^2\left(a_2^2-x^2\right)-\frac{2\lambda}{3}\left(a_2^3-x^3\right)}}}\phantom{uuuuuuuuu}\nonumber\\ 
&=&\frac{1}{\omega}\int_0^{\pi/2}{\frac{d\theta}{\sqrt{1-\frac{2}{3}\frac{\lambda|a_1|}{\omega^2}\left(\frac{1-\sin^3{\theta}}{\cos^2{\theta}}\right)}}}+\frac{1}{\omega}\int_0^{\pi/2}{\frac{d\theta}{\sqrt{1+\frac{2}{3}\frac{\lambda a_2}{\omega^2}\left(\frac{1-\sin^3{\theta}}{\cos^2{\theta}}\right)}}}\phantom{uuuuuuuuuuuuuuuuu}\nonumber\\
&=&\frac{1}{\omega}\int_0^{\pi/2}d\theta\Big[1+\frac{\lambda|a_1|}{3\omega^2}\left(\frac{1-\sin^3{\theta}}{\cos^2{\theta}}\right)+\frac{3}{8}\left(\frac{2\lambda|a_1|}{3\omega^2}\right)^2\left(\frac{1-\sin^3{\theta}}{\cos^2{\theta}}\right)^2\phantom{uuuuuuuuuuuuuuuuu}\nonumber\\ &&+1-\frac{\lambda a_2}{3\omega^2}\left(\frac{1-\sin^3{\theta}}{\cos^2{\theta}}\right)+\frac{3}{8}\left(\frac{2\lambda}{3\omega^2}\right)^2\left(\frac{1-\sin^3{\theta}}{\cos^2{\theta}}\right)^2\Big]\phantom{uuuuuuuuuuuuuuu}\label{p1,37}
\end{eqnarray}
Noting that $|a_1|$ and $a_2$ differ at $O(\lambda)$, we see immediately that the corrections to the leading order answer of $\pi$ for the integral comes at $\mathcal{O}(\lambda^2)$ and is given by
\begin{eqnarray}
\frac{T}{2} = \frac{\pi}{\omega}+\frac{\lambda^2}{\omega}\frac{4E}{9\omega^6} \int_0^{\pi/2}\left[\frac{1-\sin^3{\theta}}{\cos^2{\theta}}+\frac{3}{2}\left(\frac{1-\sin^3{\theta}}{\cos^2{\theta}}\right)^2\right] d\theta\nonumber\\
=\frac{\pi}{\omega}+\frac{\lambda^2}{\omega}\frac{4E}{9\omega^6}\left(\frac{15}{8}\pi\right)=\frac{\pi}{\omega}+\frac{\lambda^2}{\omega}\frac{5}{6}\frac{E}{\omega^6}\pi\phantom{uuuuuuuuuuuu}\label{p1,eqn38}
\end{eqnarray}
To implement the Lindstedt-Poincare scheme, we expand as before($\Omega$ is the real frequency of oscillations)
\begin{eqnarray}
x=x_0+\lambda x_1+\lambda^2 x_2+\dots \\ \label{p1,eqn39}
\Omega^2=\omega^2+\lambda\omega_1^2+\lambda^2\omega_2^2+\dots \label{p1,eqn40}
\end{eqnarray}
Inserting in Eq.~\eqref{p1,eqn29} and equating identical powers of $\lambda$ on either sides
\begin{eqnarray}
\lambda^0\colon\quad\quad\ddot{x}_0+\Omega^2 x_0 &=& 0\\ \label{p1,eqn42}
\lambda^1\colon\quad\quad\ddot{x}_1+\Omega^2 x_1 &=& -x_0^2+\omega_1^2 x_0\\ \label{p1,eqn43}
\lambda^2\colon\quad\quad\ddot{x}_2+\Omega^2 x_2 &=& -2 x_0 x_1+\omega_1^2 x_1+\omega_2^2 x_0 \label{p1,eqn44}
\end{eqnarray}
The solution of Eq.~\eqref{p1,eqn42} (initial conditions $x_0(0)=0$,\text{ }$\dot{x}_0(0)=0$) is,
\begin{equation}\label{p1,eqn45}
x_0=A\cos{\Omega t}
\end{equation}
Using this in the Eq.~\eqref{p1,eqn43} we have,
\begin{equation}\label{p1,eqn46}
\ddot{x}_1+\Omega^2 x_1 = -\frac{A^2}{2}\left(1+\cos{2\Omega t}\right)+\omega_1^2\cos{\Omega t}
\end{equation}
As before, there is a resonating term on the right hand side and removal of this requires
\begin{equation}\label{p1,eqn47}
\omega_1 =0
\end{equation}
The solution $x_1$ is now
\begin{equation}\label{p1,eqn48}
x_1=B_1\cos{\Omega t}+B_2 \sin {\Omega t}-\frac{A^2}{2\Omega^2}+\frac{A^2}{6\Omega^2}\cos{2\Omega t}
\end{equation}
Using $x_1=\dot{x}_1=0$ at $t=0$, we get $B_2=0$ and $B_1=\frac{A^2}{3\Omega^2}$. With this information, Eq.~\eqref{p1,eqn44} reads
\begin{equation}\label{p1,eqn49}
\ddot{x}_2+\Omega^2 x_2 = \frac{A^3}{3\Omega^2}\left(1+\cos{2\Omega t}\right)-\frac{A^3}{6\Omega^2}\left(\cos{3\Omega t}+\cos{\Omega t}\right)+\omega_2^2 A\cos{\Omega t}
\end{equation}
Once again the resonating terms need to be removed from the right hand side and this leads to
\begin{equation}\label{p1,eqn50}
 \omega_2^2 = -\frac{5A^2}{6\Omega^2}
\end{equation}
The frequency thus to $O(\lambda^2)$ is $\Omega=\omega-\frac{5}{12}\frac{\lambda^2A^2}{\omega^3}+\dots$ in agreement with Eq.~\eqref{p1,eqn38} and the text of Landau and Lifshitz.

A harmonic balance approach requires an expansion
\begin{equation}\label{p1,eqn51}
x=\alpha_1+\alpha_2\cos{\Omega t}+\alpha_3\cos{2\Omega t}+\dots
\end{equation}
The equation of motion becomes
\begin{eqnarray}
-\Omega^2\alpha_2\cos{\Omega t}-4\Omega^2\alpha_3\cos{2\Omega t}+\dots+\omega^2\left(\alpha_1+\alpha_2\cos{2\Omega t}\right)\phantom{uuuuuuuuuuuuuuuuuuu}\nonumber\\
+\lambda\left(\alpha_1^2+2\alpha_1\alpha_2\cos{\Omega t}+\frac{\alpha_2^2}{2}\left(1+\cos{2\Omega t}\right)+2\alpha_1\alpha_3\cos{2\Omega t}+\alpha_2\alpha_3\cos{\Omega t}+\dots\right)=0\label{p1,eqn52}
\end{eqnarray}
Setting the coefficient of each harmonic separately to zero, we get
\begin{eqnarray}
\frac{\alpha_2^2}{2}+\alpha_1^2 &=& -\frac{\omega^2}{\lambda}\alpha_1 \\ \label{p1,eqn53}
\alpha_2\left(\omega^2-\Omega^2+2\lambda\alpha_1+\lambda\alpha_3\right) &=& 0 \\ \label{p1,eqn54}
\alpha_3\left(\omega^2-4\Omega^2\right)+\frac{\lambda\alpha_2^2}{2}+2\lambda\alpha_1\alpha_3 &=& 0 \label{p1,eqn55}
\end{eqnarray}
If we ignore $\alpha_3$, as recommended in Jordan and Smith, then the perturbative solutions yields
\begin{eqnarray}
\alpha_1 &\simeq& -\frac{\lambda\alpha_2^2}{2\omega^2} \\ \label{p1,eqn56}
\text{and}\quad\Omega^2 &=& \omega^2+2\lambda\alpha_1 = \omega^2-\frac{\lambda^2\alpha_2^2}{\omega^2} \label{p1,eqn57}
\end{eqnarray}
in agreement with the result quoted there and in disagreement with the $\mathcal{O}(\lambda^2)$ term of Landau and Lifshitz and the direct result obtained in Eq.~\eqref{p1,eqn38}. Keeping the $\alpha_3$ term,
\begin{equation}\label{p1,eqn58}
\alpha_3\simeq-\frac{\lambda\alpha_2^2}{2\left(\omega^2-4\Omega^2\right)}=\frac{\lambda\alpha_2^2}{6\omega^2}+O(\lambda^2)
\end{equation}
Using this in Eq.~\eqref{p1,eqn54} we finally get
\begin{equation}\label{p1,eqn59}
\Omega^2 = \omega^2 - \frac{5}{6}\frac{\lambda^2\alpha_2^2}{\omega^2}
\end{equation}
which is the correct answer at $\mathcal{O}(\lambda^2)$.


\section{LOTKA-VOLTERRA MODEL}
The predator-prey system is a two variable dynamical system with $x$ standing for the number of prey at time $t$ and $y$ the number of predators. Left to themselves, the number of prey grows(rabbit feeding on grass) while the number of predators decreases (foxes left with no food). An interaction between the two species causes the number of prey to fall and predators to increase. An interaction between the two species causes the number of preys to fall and the predators to increase. The interaction effect is jointly proportional to the number of prey and predators. These considerations led to the Lotka-Volterra model
\begin{eqnarray}\label{p1,eqn60}
\begin{split}
\frac{dx}{dt} &=& x-xy\phantom{u}\\ 
\frac{dy}{dt} &=& -y+xy
\end{split}
\end{eqnarray}
The model has two fixed points $x=y=0$ which is unstable and $x=y=1$which is a center, i.e. small deviations from it execute a periodic motion. In reality there is a periodic orbit about the point $(1,1)$ and it is the time period of that orbit which we would like to determine.

The first step is to introduce variable $x_1,x_2$ centered around $(1,1)$ where $x_1=x-1$ and $x_2=y-1$
and in terms of these variables, the equation of motion become
\begin{eqnarray}\label{p1,eqn62}
\begin{split}
\dot{x}_1 &=& -x_2-x_1x_2\\
\dot{x}_2 &=& x_1 + x_1 x_2\phantom{u}
\end{split}
\end{eqnarray}
It is clear from the above that for small $x_1,x_2$ the linearized system $\dot{x_1}=-x_2$ and $\dot{x_2}=x_1$ corresponds to simple harmonic motion with frequency unity. The nonlinear terms make the time period, amplitude dependent. This is what we set out to find. To do so we introduce a parameter $\lambda$ in front of the nonlinear term and the write Eq.~\eqref{p1,eqn62} as
\begin{eqnarray}\label{p1,eqn63}
\begin{split}
\dot{x}_1 &=& -x_2-\lambda x_1x_2\\
\dot{x}_2 &=& x_1+\lambda x_1x_2\phantom{u}
\end{split}
\end{eqnarray}
For $\lambda = 0$ the dynamics is that of a simple harmonic oscillator with $x_1=A_1\cos{t},\text{ }x_2=A_2\cos{t}$ where $A_1$ and $A_2$ are the initial values of $x_1$ and $x_2$.
For $\lambda\neq 0$, we explore the possibility of existence of this periodic solution. The frequency of motion will no longer be unity but some frequency $\Omega$ which needs to be determine. We anticipate for small $\lambda$
\begin{equation}\label{p1,eqn64}
\Omega=1+\lambda\omega_1+\lambda^2\omega_2+\dots
\end{equation}
while
\begin{equation}\label{p1,eqn65}
x_{1,2}=x_{10,20}+\lambda x_{11,21}+\lambda^2 x_{12,22}+\dots
\end{equation}
We rewrite the Eq.~\eqref{p1,eqn63} as
\begin{eqnarray}\label{p1,eqn66}
\begin{split}
\dot{x}_1 &=& -\Omega x_2-\lambda x_1 x_2+\left(\lambda\omega_1+\lambda^2\omega_2+\dots\right)x_2\\
\dot{x}_2 &=& \Omega x_1+\lambda x_1 x_2+\left(\lambda\omega_1-\lambda^2\omega_2+\dots\right)x_2\phantom{u}
\end{split}
\end{eqnarray}
We now expand $x_{1,2}$ as in Eq.~\eqref{p1,eqn65} and insert the series in Eq.~\eqref{p1,eqn66}. Collecting and equating identical powers of $\lambda$ on either side from the resulting equation, we get
\begin{eqnarray}\label{p1,eqn68}
\begin{split}
\lambda^0\colon\quad\quad\quad\quad\dot{x}_{10}+\Omega x_{20} &=& 0\phantom{\left(x_{10}x_{21}+x_{20}x_{11}\right)+\omega_1x_{21}+\omega_2x_{20}}\\
\dot{x}_{20}-\Omega x_{10} &=& 0\phantom{\left(x_{10}x_{21}+x_{20}x_{11}\right)+\omega_1x_{21}+\omega_2x_{20}}\\
\end{split}
\end{eqnarray}
\begin{eqnarray}\label{p1,eqn69}
\begin{split}
\lambda^1\colon\quad\quad\quad\quad\dot{x}_{11}+\Omega x_{21} &=& -x_{10}x_{20}+\omega_1x_{20}\phantom{+\omega_1x_{21}+\omega_2x_{20}uuu}\\ 
\dot{x}_{21}-\Omega x_{11} &=& x_{10}x_{20}-\omega_1x_{10}\phantom{+\omega_1x_{21}+\omega_2x_{20}uuu}\\ 
\end{split}
\end{eqnarray}
\begin{eqnarray}\label{p1,eqn70}
\begin{split}
\lambda^2\colon\quad\quad\quad\quad\dot{x}_{12}+\Omega x_{22} = -x_{10}x_{21}-x_{20}x_{11}+\omega_1x_{21}+\omega_2x_{20}\phantom{uuu}\\
\dot{x}_{22}-\Omega x_{12} = x_{10}x_{21}+x_{20}x_{11}-\omega_1x_{11}-\omega_2x_{10}\phantom{uuuuu}
\end{split}
\end{eqnarray}
Clearly at $\mathcal{O}(\lambda^0)$ the solution is, $x_{10}=A_1\cos{\Omega t}+A_2\sin{\Omega t},x_{20}=A_1\sin{\Omega t}-A_2\cos{\Omega t}$ where $A_1$ and $A_2$ are taken to be the initial values of $x_1$ and $x_2$. At $\mathcal{O}(\lambda^1)$, we have
\begin{subequations}\label{p1,eqn71}
\begin{equation}\label{p1,eqn71a}
\dot{x}_{11}+\Omega x_{21} = -\frac{A_1^2-A_2^2}{2}\sin{2\Omega t}
+\frac{A_1A_2}{2}\cos{2\Omega t}+\omega_1\left(A_1\sin{\Omega t}-A_2\cos{\Omega t}\right)
\end{equation}
\begin{equation}\label{p1,eqn71b}
\dot{x}_{21}-\Omega x_{21} = \frac{A_1^2-A_2^2}{2}\sin{2\Omega t}
-\frac{A_1A_2}{2}\cos{2\Omega t}+\omega_1\left(A_1\cos{\Omega t}+A_2\sin{\Omega t}\right) 
\end{equation}
\end{subequations}
The homogeneous equation has the solution of the form $A\cos{\Omega t}+B\sin{\Omega t}$ which shows that the $\cos{\Omega t}$ and $\sin{\Omega t}$ in Eq.~\eqref{p1,eqn71} will cause the system to resonate. removal of this term requires $\omega_1=0$. Now from Eq.~\eqref{p1,eqn71}
\begin{equation}\label{p1,eqn72}
\ddot{x}_{11}+\Omega^2 x_{11}=\Omega\left(A_1A_2-\left(A_1^2-A_2^2\right)\right)\cos{2\Omega t}-\Omega\left(2A_1A_2+\frac{1}{2}\left(A_1^2-A_2^2\right)\right)\sin{2\Omega t}
\end{equation}
yielding 
\begin{equation}\label{p1,eqn73}
x_{11}=A\cos{\Omega t}+B\sin{\Omega t}-\frac{1}{3\Omega}\left[A_1A_2-\left(A_1^2-A_2^2\right)\right]\cos{2\Omega t}+\frac{1}{6\Omega}\left[\left(A_1^2-A_2^2\right)+4 A_1A_2\right]\sin{2\Omega t}
\end{equation}

At $t=0,x_{11}=0$ which sets $A=\frac{1}{3\Omega}\left[A_1A_2-\left(A_1^2-A_2^2\right)\right]$. From Eq.~\eqref{p1,eqn71} we have,
\begin{eqnarray}
\Omega x_{21}=-B\Omega\cos{\Omega t}-\frac{1}{3\Omega}\left[A_1A_2-\left(A_1^2-A_2^2\right)\right]\sin{\Omega t}\phantom
{uuuuuuuuuuuuuuuuuuuuuu}\nonumber\\-\frac{1}{3\Omega}\left[4A_1A_2+\left(A_1^2-A_2^2\right)\right]\cos{\Omega t}-\frac{\left(A_1^2-A_2^2\right)}{2}\sin{\Omega t}+A_1A_2\cos{\Omega t}\phantom{\Omega\cos{\Omega t}}\label{p1,eqn74}
\end{eqnarray}
At $t=0,x_{11}=0$ leads to $B=-\frac{1}{3\Omega}\left[A_1A_2+\left(A_1^2-A_2^2\right)\right]$. So Finally we have
\begin{subequations}\label{p1,eqn75}
\begin{equation}\label{p1,eqn75a}
 x_{11}=A\cos{\Omega t}+B\sin{\Omega t}+C\cos{2\Omega t}+D\sin{2\Omega t}
\end{equation}
\begin{equation}\label{p1,eqn75b}
 x_{21}=A\cos{\Omega t}-B\sin{\Omega t}+E\cos{2\Omega t}+F\sin{2\Omega t}
\end{equation}
\end{subequations}
where,
\begin{eqnarray}\label{p1,eqn76}
\begin{split}
A&=&\frac{1}{3\Omega}\left[A_1A_2-\left(A_1^2-A_2^2\right)\right],\quad\quad\\
C&=&-\frac{1}{3\Omega}\left[A_1A_2-\left(A_1^2-A_2^2\right)\right],\quad\quad\\
F&=&\frac{1}{6\Omega}\left[-4A_1A_2+\left(A_1^2-A_2^2\right)\right],\quad\quad
\end{split}
\begin{split}
B&=&-\frac{1}{3\Omega}\left[A_1A_2+\left(A_1^2-A_2^2\right)\right]\\
D&=&\frac{1}{6\Omega}\left[4A_1A_2+\left(A_1^2-A_2^2\right)\right]\\ 
E&=&-\frac{1}{3\Omega}\left[A_1A_2+\left(A_1^2-A_2^2\right)\right]
\end{split}
\end{eqnarray}
Now at $\mathcal{O}(\lambda^2)$ we have the system of equations
\begin{eqnarray}\label{p1,eqn77}
\begin{split}
\dot{x}_{12}+\Omega x_{22} &=& -\left(x_{10}x_{21}+x_{20}x_{11}\right)+\omega_2x_{20}\\ 
\dot{x}_{22}-\Omega x_{12} &=& \left(x_{10}x_{21}+x_{20}x_{11}\right)-\omega_2x_{10}
\end{split}
\end{eqnarray}
We first compute the term $\left(x_{10}x_{21}+x_{20}x_{11}\right)$. Using results obtained for $x_{10},x_{20},,x_{11}x_{21}$ we get

\begin{eqnarray}
x_{10}x_{21}+x_{20}x_{11}=\left(A_1\cos{\Omega t}+A_2\sin{\Omega t}\right)\left(A\cos{\Omega t}+B\sin{\Omega t}+C\cos{2\Omega t}+D\sin{2\Omega t}\right)\phantom{uuuu}\nonumber\\+\left(A_1\sin{\Omega t}-A_2\cos{\Omega t}\right)\left(A\cos{\Omega t}-B\sin{\Omega t}+E\cos{2\Omega t}+F\sin{2\Omega t}\right)\phantom{uuu}\nonumber\\
=\left[\text{higher harmonics}\right]+\frac{1}{2}\left(A_1E-A_2C+A_1D+A_2F\right)\cos{\Omega t}\phantom{uuuuuuuuuuuu}\nonumber\\+\frac{1}{2}\left(A_2E+A_1C-A_1F+A_2D\right)\sin{\Omega t}\phantom{uuuuuuuuuuuuuuuuuuuuuuuuuuuu}\label{p1,eqn78}
\end{eqnarray}

We need not pay heed to coefficients of the higher harmonics as we just need to look at terms that will cause the system to resonate. Such terms must be put to zero from physical considerations. From Eq.~\eqref{p1,eqn77} we get
\begin{eqnarray}
\ddot{x}_{12}+\Omega^2_{12}
&=&-\Omega\left(x_{10}x_{21}+x_{20}x_{11}\right)+\Omega\omega_2x_{10}-\frac{\partial}{\partial t}\left(x_{10}x_{21}+x_{20}x_{11}\right)+\omega_2\dot{x}_{20}\phantom{uuuuuuuuu}\nonumber\\  &=&-\frac{\Omega}{2}\left(A_1E-A_2C+A_1D+A_2F+A_2E+A_1C-A_1F+A_2D\right)\cos{\Omega t}\phantom{uuuuu}\nonumber\\ &&-\frac{\Omega}{2}\left(A_2E+A_1C-A_1F+A_2D-A_1E+A_2C-A_1D-A_2F\right)\sin{\Omega t}\phantom{uuuuu}\nonumber\\ &&+2\Omega\omega_2\left(A_1\cos{\omega t}+A_2\sin{\Omega t}\right)+\frac{\Omega}{2}\left(A_1E-A_2C+A_1D+A_2F\right)\sin{\Omega t}\phantom{uuuuu}\label{p1,eqn79}
\end{eqnarray}
Now to get rid of the resonance creating terms $\cos{\Omega t}$ and $\sin{\Omega t}$ we must take their coefficients to be zero.Thus we get
\begin{equation}\label{p1,eqn80}
2\omega_2A_1 =\frac{1}{2}[-A_1E+A_2C-A_1D-A_2F-A_2E-A_1C+A_1F-A_2D]
\end{equation}
\begin{equation}\label{p1,eqn81}
2\omega_2A_2 =\frac{1}{2}[-A_2E-A_1C+A_1F-A_2D+A_1E-A_2C+A_1D+A_2F]
\end{equation}
From Eqs.~\eqref{p1,eqn80} and ~\eqref{p1,eqn81} we obtain
\begin{equation}\label{p1,eqn82}
2\omega_2\left(A_1+A_2\right)=A_1\left(F-C\right)-A_2\left(E+D\right)
\end{equation}
Now using the expressions for $A_1, A_2, C,D,E, \text{and } F$ we get
\begin{equation}\label{p1,eqn83}
2\omega_2\left(A_1+A_2\right)=\frac{A_2^2-A_1^2}{6\Omega}\left(A_1-A_2\right)-\frac{A_1A_2}{3\Omega}\left(A_1+A_2\right)
\end{equation}
Finally we have the expression for $\omega_2$
\begin{equation}\label{p1,eqn84}
\omega_2=-\frac{A_1^2+A_2^2}{12\Omega}
\end{equation}
The perturbative result for $\Omega$ up to $O(\lambda^2)$, after setting $\lambda =1$ is
\begin{equation}\label{p1,eqn85}
\Omega=1-\frac{|A|^2}{12\Omega}
\end{equation}
where $A$ is the amplitude of the limit cycle. If $x_0$ and $y_0$ be the initial values of $x$ and $y$, then we can write
\begin{equation}\label{p1,eqn86}
\Omega=1-\frac{\left(x_0-1\right)^2+\left(y_0-1\right)^2}{12}
\end{equation}
\begin{figure}
\centering
\includegraphics[scale=.3,angle=270]{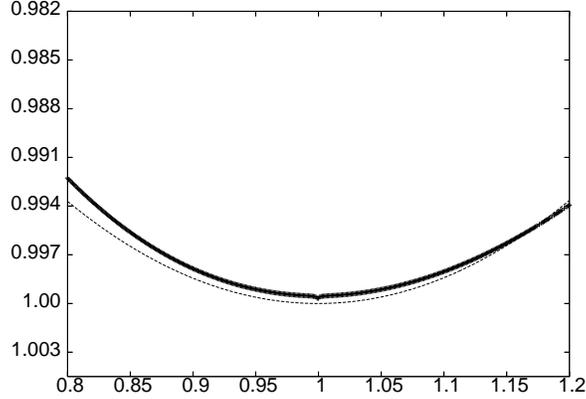}
\caption{Initial number of population density modifies its periodicity. Initial potulation density (x=y) is plotted with corresponding frequency ($\omega$) of its oscillation. Dotted line: \textit{Theoretical prediction}. Solid Line: \textit{Numerical result}}
\label{Fig[1]}
\end{figure}
We have checked this result numerically. The results are shown in Fig[1]. The good agreement between the computed frequencies and obtained from Eq.~\eqref{p1,eqn85} is apparent.

\section{EQUIVALENT LINEARIZATION AND CONCLUSION}
In this section, we conclude by explaining the equivalent linearization process which spans the entire range from low to high expansion parameter. As will be apparent in the immensely useful process the input coming from the low order perturbation theory is vital.

The technique of equivalent linearization proceeds by replacing the non-linear term $x^3$ by a linear term $\alpha\langle x^2\rangle x$, where $\alpha$ is a number of $\mathcal{O}(1)$ and $\langle x^2\rangle$ is the average of $x^2$ over one cycle. We now simply write Eq.~\eqref{p1,eqn1} as the equation of motion of a simple harmonic oscillator
\begin{equation}\label{p1,eqn23}
\ddot{x}+\Omega^2 x=0\quad\quad\text{where}\quad\Omega^2=\omega^2\lambda\alpha\langle x^2\rangle
\end{equation}
The first question is how do we fix the value of $\alpha$? In terms of the potential, we have replaced $\frac{\lambda}{4}x^4$ by $\frac{\lambda}{2}\alpha\langle x^2 \rangle x^2$ and if we demand that this statement be true at least on an average, then we would require
\begin{equation}\label{p1,eqn24}
\langle x^4 \rangle = 2\alpha\langle x^2 \rangle^2
\end{equation}
and using periodic solution $x=A_0\cos{\Omega t}$, this leads to $\alpha=3/4$. Writing $\langle x^2 \rangle = \frac{A_0^2}{2}$, Eq.~\eqref{p1,eqn23} shows $\Omega^2=\omega^2+\frac{3}{8}\lambda A_0^2$ or
\begin{equation}\label{p1,eqn25}
\Omega=\omega\left(1+\frac{3}{16}\lambda\frac{A_0^2}{\omega^2}\right)
\end{equation}
which differ from the Lindstedt-Poincare answer at $\mathcal{O}(\lambda)$ by a factor of two. In using the equivalent linearization technique, it is best to use it in conjunction with the Poincare Lindstedt perturbation theory to fix the parameters of the linearization scheme. In this particular case the agreement with the correct perturbation theory requires fixing of $\alpha$ as $\alpha= 3/2$.
The strength of the equivalent linearization lies elsewhere. With the equivalent simple harmonic oscillator of Eq.~\eqref{p1,eqn23} the integral of motion is $\frac{1}{2}\dot{x}^2+\frac{1}{2}\Omega^2 x^2$ which is recognized as the energy of the oscillator. The energy is a physical quantity which is fixed by the initial condition and the primary quantity for the dynamics as opposed to the amplitude. For the simple harmonic oscillator $\frac{1}{2}\Omega^2 A^2 = E$, where $A$ is the amplitude while for the original oscillator $E=\frac{1}{2}\omega^2 a^2+\frac{\lambda a^4}{4}$, with $a$ being the corresponding amplitude. Returning to our equivalent oscillator now, we write $\langle x^2\rangle=\frac{A^2}{2}=\frac{E}{\Omega^2}$ and using Eq.~\eqref{p1,eqn23} with $\alpha=3/2$, find
\begin{equation}\label{p1,eqn26}
\Omega^2=\omega^2+\frac{3}{2}\frac{\lambda E}{\Omega^2}
\end{equation}
with the result
\begin{equation}\label{p1,eqn27}
\Omega^2=\frac{1}{2}\left[\omega^2+\sqrt{\omega^4+6\lambda E}\right]
\end{equation}
Expanding in powers of $\lambda$, this gives the correct answer to $\mathcal{O}(\lambda)$ as it should but goes further than any perturbation theory by giving a meaningful large $\lambda$ limit. For $\lambda E/\omega^4\gg1$, Eq.~\eqref{p1,eqn27} shows
\begin{equation}\label{p1,eqn28}
\Omega\simeq\left(\frac{3}{2}\right)^{1/4}\left(\lambda E\right)^{1/4}
\end{equation}
which is to be compared with the exact answer of $\Omega\simeq\left[\sqrt{2}\pi/\Gamma(\frac{1}{4})\right]\left(\lambda E\right)^{1/4}$. The two prefactors differ by a meager $10\%$. The entire range of $\lambda$ can be handled by Eq.~\eqref{p1,eqn27}), wherein lies the strength of equivalent linearization. It is important to note that Lindstedt-Poincare technique provides a necessary input for the success of equivalent linearization.

\end{document}